\documentclass[journal,twoside]{IEEEtran}
\usepackage[noadjust,compress]{cite}
\usepackage{graphicx}
\usepackage[pdftex]{hyperref}
\usepackage[cmex10]{amsmath}
\usepackage{amssymb}
\usepackage{color}
\usepackage[utf8]{inputenc}

\usepackage{todonotes}

\newcommand{\prob}[1]{\mbox{Prob}\left\{#1\right\}}
\newcommand{\E}[1]{\mbox{$\mathbb{E}[#1]$}}
\newcommand{\ie}{\textit{i}.\textit{e}.}
\newcommand{\eg}{\textit{e}.\textit{g}.}
\newcommand{\etal}{\textit{et al}.}
\newcommand{\cf}{\textit{cf}.}

\title{Capacity Value of Solar Power and Other Variable Generation}

\author{S.\ Awara, M.\ Lynch, S.\ Pfenninger, K.\ Schell, R.\ Sioshansi, I.\ Staffell, N.\ Samaan, S.H.\ Tindemans, A.L.\ Wilson, S.\ Zachary, H.\ Zareipour, C.J.\ Dent}

\IEEEspecialpapernotice{Report of the IEEE PES Task Force on Capacity Value of Solar Power}

\begin{document}
\maketitle

\begin{abstract}
%\ils{Perhaps 1 sentence of context to preface this?  The capacity value of power generation technologies is critically important as it quantifies xxxxx...}
%\cjd{From PMAPS paper}
This paper reviews methods that are used for adequacy risk assessment considering solar power and for assessment of the capacity value of solar power. The properties of solar power are described as seen from the perspective of the power-system operator, comparing differences in energy availability and capacity factors with those of wind power. Methodologies for risk calculations considering variable generation are surveyed, including the probability background, statistical-estimation approaches, and capacity-value metrics. Issues in incorporating variable generation in capacity markets are described, followed by a review of applied studies considering solar power. Finally, recommendations for further research are presented.
\end{abstract}


\section{Introduction}
\label{sec:intro}
%\cjd{This text is the intro from the PMAPS paper \cite{dent2016a}} 
\IEEEPARstart{A}{}key issue for power-system planning is the contribution of renewable and other emerging energy resources to meeting demand reliably. Mechanical failures, planned maintenance, or lack of generating resource in real-time may leave a system with insufficient capacity to meet load---requiring load curtailment. The contribution of a resource to serving demand reliably is measured typically by estimating capacity-value metrics, defined through the effect that its addition to the system has on the calculated risk of load-curtailment events. The issue of real-time resource availability is particularly salient with renewable resources, as their output is governed by uncontrollable weather conditions.

An IEEE Task Force focused on techniques for estimating the capacity value of wind power published a survey on that technology \cite{Keane2011a}. This new paper has a similar purpose of surveying methods for estimating the capacity value of solar power and recent activity applicable to both wind and solar. We place strong emphasis on critical review of modelling methodology, particularly with respect to capacity markets and statistical modelling, which distinguishes our review work from related publications \cite{Keane2011a,Soder2019,doorman2016}. The paper builds on earlier Task Force papers which concentrate more specifically on solar power \cite{duignan2012a,dent2016a}---while the high-level topics covered in this new paper are broadly similar to those in a previous conference paper \cite{dent2016a}, the material is revised entirely for this as the Task Force's final report apart from Sections~\ref{sec:methodology:probability}--\ref{sec:methodology:metrics} (these cover the essentials of the relevant probabilistic and statistical modelling, where the Task Force's thinking has evolved less rapidly.) Throughout the Task Force's activity, there is particular emphasis on matters of solar-resource assessment (with which the power-system community may be less familiar as compared to wind). In the solar-specific sections, we focus on photovoltaic (PV) solar rather than concentrating solar power (CSP). CSP has intrinsic energy-storage capability \cite{pfenninger2014a,madaeni2013b}, providing some control of co-incidence of output with high demands. This characteristic of CSP makes relevant modelling approaches fundamentally different from PV. A brief discussion regarding the interaction between solar power and co-located energy storage, which is applicable to CSP, is given in Section~\ref{sec:methodology:hybrid}.  

This paper addresses four major issues that are related to solar power. First, Section~\ref{sec:resource} discusses key properties and assessment of solar resource. Solar availability features unique spatial and temporal correlations, which are modified by design considerations such as panel orientation and the inclusion of sun-tracking systems or energy storage. Section~\ref{sec:methodology} provides a detailed discussion of the statistical methods that are used for adequacy-index and capacity-value estimation---much of which applies equally to other variable generation (VG) technologies, as well as solar. We highlight the importance of capturing statistical relationships between renewable resource and demand, and consequences of limited data. It discusses also relevant theory associated with capacity markets. %\cjd{Is this a major theme in the paper: Another topic of interest is that of using statistical models to determine what properties of the solar pattern affect the capacity value of solar. Such insights can provide useful screening tools to determine how to design solar systems and policies to maximize capacity value.}  
Section~\ref{sec:survey} surveys recent capacity value studies and practice in the industrial and research literature, emphasising consequences of different methodology choices. Finally, Section~\ref{sec:concl} concludes and discusses key research needs in this area.

\section{PV-Resource Assessment}
\label{sec:resource}
\begin{figure}
\centering
\includegraphics[width=\columnwidth]{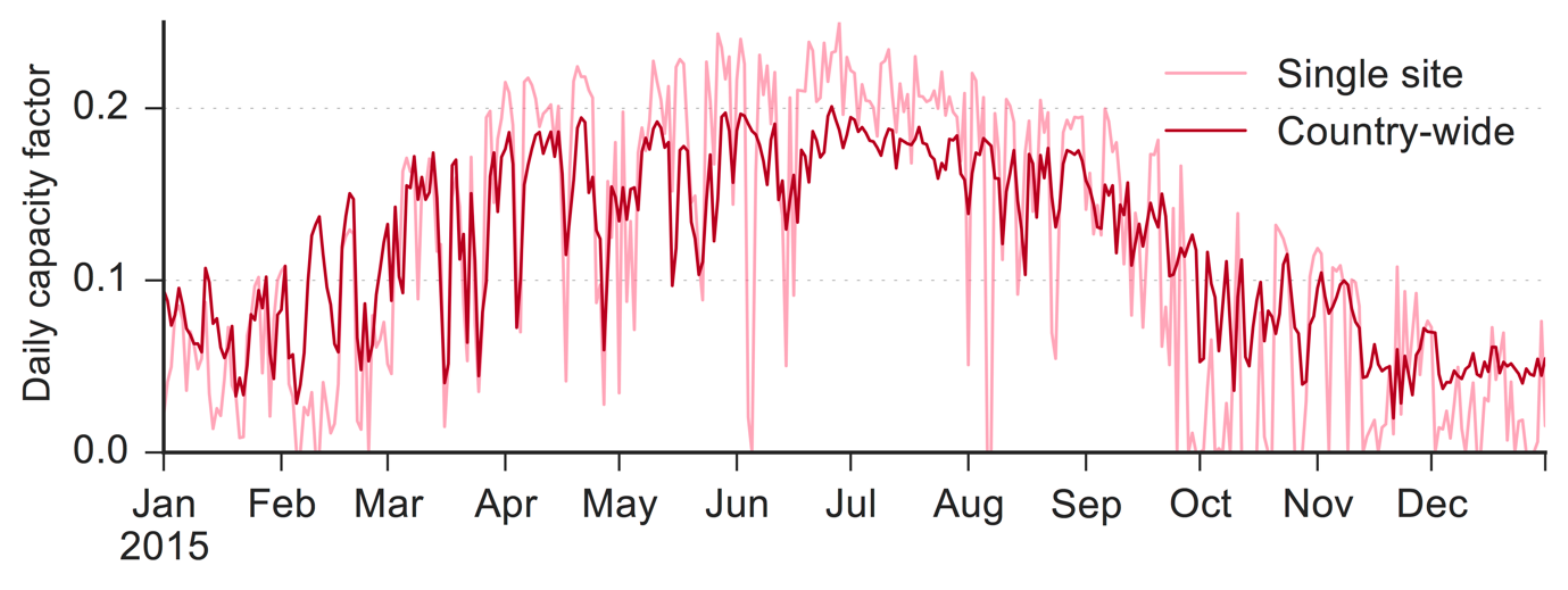}
\caption{Daily-average capacity factor for a single PV system near Milan and for PV deployed across Italy during~$2015$. Single-site and Italy-wide data are from PVOutput and Terna, respectively.}
\label{fig:resource:italy}
\end{figure}

Surface solar irradiance follows predictable diurnal and seasonal cycles. However, solar irradiance can be difficult to model and forecast, due to cloud cover and other meteorological effects. The recent emergence of PV and its distributed nature make reliable long-term output data rare, forcing reliance on modeled PV-generation data \cite{pfenninger2016a}. Weather variability occurs at different temporal and spatial scales, from clouds moving across individual panels (seconds to minutes \cite{gami2017a}) to weather fronts moving over a region (hours to days \cite{Jewell1987a}) to multi-day regimes that dictate continental-scale weather patterns \cite{grams2017a}. Fig.~\ref{fig:resource:italy} demonstrates the variability in PV output at a single location over short timescales and that this variability is reduced if many PV systems over a wide area are aggregated.

\begin{figure}
\centering
\includegraphics[width=\columnwidth]{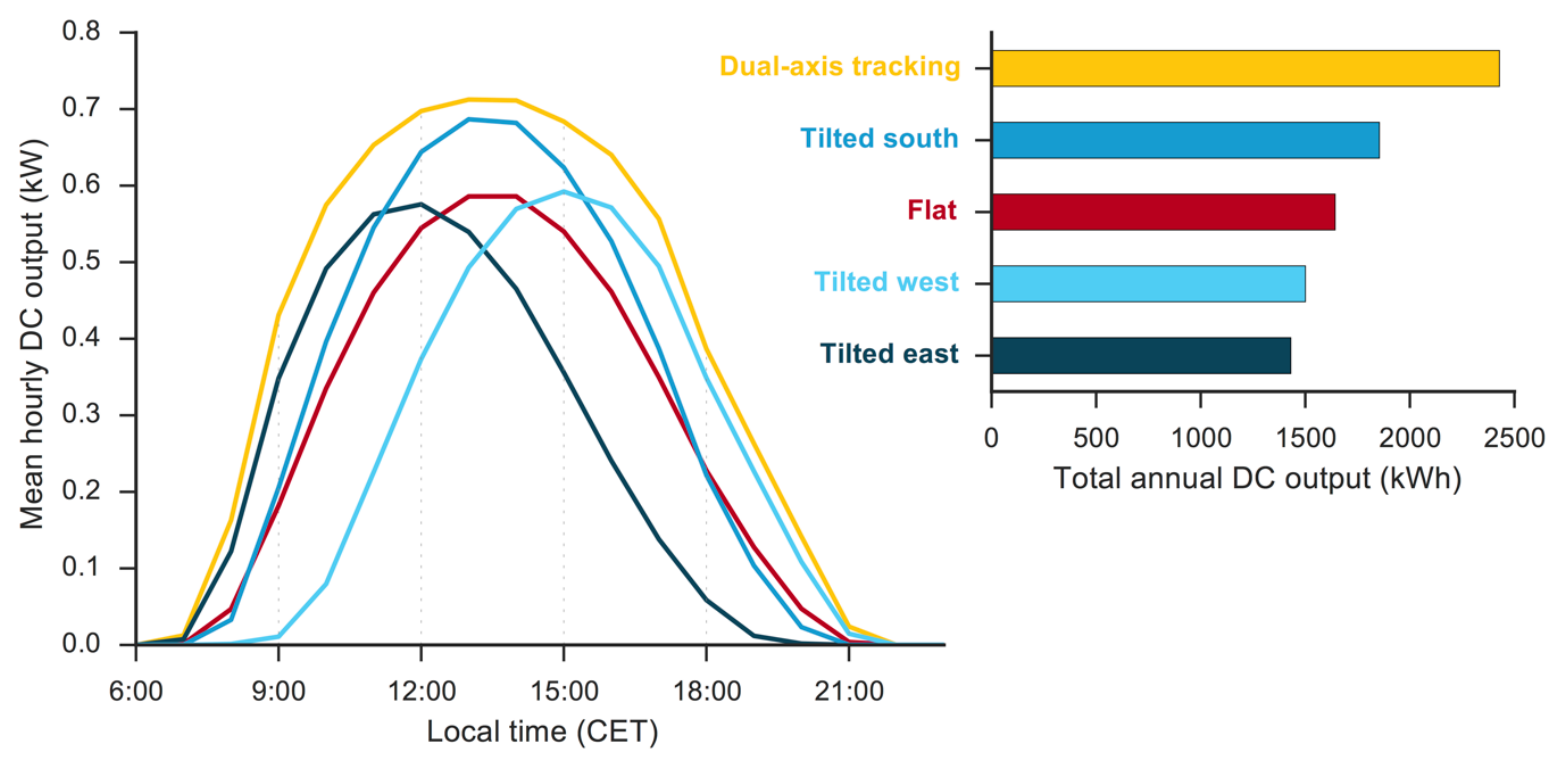}
\caption{Simulated power generation for a $1$-kW system installed in Jaen, Spain, averaged hourly over all days of~$2015$ (left) and summed over the entire year (right). The tilted systems are installed at $35$-degree tilt angle, facing exactly south, east, or west. Data from \url{https://www.renewables.ninja}.}
\label{fig:resource:orientation}
\end{figure}

Modelling PV power output accurately is hampered by the difficulty of estimating solar irradiance \cite{coimbra2013a}, especially due to cloud cover. Aerosols and other atmospheric particles scatter incoming light even with clear skies \cite{schroedter-homscheidt2013a}, affecting the productivity of concentrating technologies (CSP and concentrating PV) and, to a lesser degree, PV. Moreover, deposition of aerosols and particles on panels affects productivity \cite{Micheli2017a}. Output depends also on many secondary parameters: the PV technology that is used, tilt and azimuth angles, whether panels are fixed or have tracking systems, module temperature \cite{huld2010a}, and panel shading as a function of sun angle \cite{huld2008a}. Fig.~\ref{fig:resource:orientation} illustrates the impact of orientation on PV output, using data for Jaen, Spain. Other weather variables play a role as well: the severity of soiling is mediated by rainfall \cite{kimber2006a} and snow can cover panels (reducing output) and reflect sunlight off the ground (increasing output) \cite{Ryberg2015a}. Finally, a PV system's inverter determines AC power output, with an efficiency that depends on utilization (power level and input voltage) and operating temperature \cite{boyson2007a}. It is common for inverters to be undersized relative to peak DC output of a panel, giving flattened power-output peaks. While this affects summer peak output in particular, snow affects winter peaks. PV output during both summer and winter peaks contribute to the capacity value of a PV system.

\subsection{Calculating Power Output}
A key challenge in modeling overall system performance is obtaining accurate irradiance data. Several methods exists to convert irradiance to DC power output from PV panels. Common approaches are empirical models, which are parameterized from manufacturer datasheets, and experimental data \cite{soto2006a,huld2010a}. The two primary weather inputs---module irradiance and temperature---are modified by the secondary parameters that are described above, requiring assumptions (\eg, on panel orientation) or additional data (\eg, aerosol optical depth or snowfall volume). These secondary parameters are of critical importance for the diurnal profile of PV generation, which, in turn, is relevant for its capacity contribution. The impacts of these secondary parameters are illustrated in Fig.~\ref{fig:resource:orientation}.

The sun's average power output (the solar constant) and inclination are fundamental values. Thus, libraries such as PVLIB \cite{holmgren2015a} can estimate overall power output over a typical meteorological year (TMY) easily. TMY data provide synthetic hourly power outputs, which are sufficient for many types of analyses \cite{Janjai2009a,Remund2010a}. The sufficiency of TMY data stem, in part, from national-scale solar capacity factors having less internannual variability compared to those for wind (\eg, $\pm 0.3$\% in Europe versus $\pm 1.5$\% for wind \cite{Staffell2016b,pfenninger2016a}). However, 
%it is necessary to rely on 
the use of TMY data requires correct depiction of the temporal and spatial dependency of PV generation under real weather conditions and preserving correlations with temperature, demand, and wind \cite{fattori2017a}.

\subsection{Sources of Irradiance and Weather Data}
There are three primary sources of data: ground-based measurements, satellite imagery, and meteorological reanalyses.

Ground-station data are best for accuracy and high temporal resolution. However, freely available data are limited and of mixed quality, suffering from missing data, measurement errors, and time aggregation. Data are available from Baseline Surface Radiation Network (BSRN) \cite{pangea2017a}, Global Energy Balance Archive (GEBA) \cite{wild2013a}, Surface Radiation Budget (SURFRAD) \cite{NOAA2017b}, Southern African Universities Radiometric Network (SAURAN) \cite{brooks2015a}, and some national weather services.

Geostationary weather satellites cover specific regions and provide half-hourly images which can be processed to derive direct and diffuse surface irradiance \cite{rigollier2004a}. Meteosat covers Europe, northern Africa, and parts of Asia, with free data available through Satellite Application Facility on Climate Monitoring (CM-SAF) \cite{kothe2017a} and Copernicus Atmosphere Monitoring Service (CAMS) \cite{Copernicus2017c}. Geostationary Operational Environmental Satellite (GOES) covers the Americas \cite{NOAA2017d}, but no equivalent data provider exists. Prospective users of GOES must process imagery themselves or use derived products, such as National Solar Radiation Data Base (NSRDB) \cite{NREL2017e}. While satellite data are considered state-of-the-art (due to high spatial resolution), they suffer from extensive periods of missing data and do not provide global coverage yet \cite{pfenninger2016a}.

Reanalyses are more consistent across space and time and provide global coverage, created by assimilating historical meteorological measurements into a numerical weather-prediction model \cite{dee2011a}. As such, reanalyses generate internally-consistent pictures of the state of the global atmosphere. Thus, reanalyses are gaining traction in simulating wind resources \cite{Staffell2016b,drew2015a,aparicio2017a}. However, spatial resolution is coarse, typically \textit{via} a $20$-km to $100$-km square grid \cite{dee2011a}. Moreover, reanalyses' focus on three-dimensional atmospheric flow means that solar irradiance so far has not been a primary consideration. Nevertheless, with appropriate bias correction, reanalyses can provide accurate PV-output simulations \cite{pfenninger2016a}. Recently, several turnkey services have launched which provide freely available PV (and wind) simulations based on reanalysis data, including from National Renewable Energy Laboratory \cite{NREL2017e}, European Climatic Energy Mixes Demonstrator \cite{EC2017f}, Photovoltaic Geographical Information System \cite{JRC2017g}, Joint Research Centre's European Meteorological derived HIgh resolution RES generation dataset (covering Europe) \cite{aparicio2017a}, and the Renewables.ninja web platform (which offers simulations that are based on CM-SAF) \cite{pfenninger2016a,Staffell2016b}.

\subsection{Measured Power-Output Data}
Metered data from individual PV systems are an alternative to simulation. These are more challenging to obtain than for other generation technologies, due to the small and distributed nature of PV. For example, there are $1.4$~million PV systems in Australia \cite{APVI2017h}, compared to less than $300$~generators registered in National Electricity Market \cite{AEMO2016a}. The lack of metered data poses a challenge to system operators, for which PV output is visible only as a reduction of demand \cite{NG2012a}.

Early government-funded field trials produced metered output data from small numbers of PV panels (\eg, $229$~systems in United Kingdom from between~$2002$ and~$2006$ \cite{munzinger2006a}). Such datasets are becoming increasingly common, with some providing comprehensive real-time updates, for example from Australian Photovoltaic Institute ($6000$~systems in Australia \cite{APVI2017h}) and Sheffield Solar ($1700$~systems in United Kingdom \cite{SS2017i}). These rely on the proliferation of web-enabled inverters, which can upload data with high temporal resolution (\eg, five-minute) data to online aggregator services.

System operators in many regions include PV output as part of their public data now. These data must be estimated, often by combining bottom-up approaches that are listed above with top-down statistical estimation \cite{SS2017i}, as operators cannot meter every PV system in a country.

\subsection{Future Improvements}
Many methodologies, including cloud imagery, physical climate models, and machine learning, are employed to improve solar-power modeling \cite{Pedro2012b,bessa2016a,coimbra2013a}. No single technique appears to be dominant for all applications. However, hybrid or ensemble machine-learning models appear to offer better accuracy than other techniques \cite{voyant2017a}. With improved models, important data issues remain: averaging data to hourly or lower resolutions, PV generation modeled inaccurately, and errors in electricity-demand data contribute uncertainty in PV capacity values \cite{gami2017a}. Even seemingly small systematic errors (\eg, a $30$-minute shift in some modelled data) can have a large impact on capacity-value estimates if they affect the relative timing of peak PV generation and demand \cite{gami2017a}.

From a decision-analytic perspective, there is also a need to build statistical error models for the relationship between resource datasets and real-world analogues, \ie, going beyond improved central estimates of historic resource. 
%for instance how reanalysis or other data represents historic meteorological conditions; the process of converting met data to power outputs; and the location and orientation of solar capacity.

Improvements in the data and modeling for solar-power prediction brings real benefits for system planning and operations, \eg, the California system must handle extensive over-generation of solar power, with system-wide curtailment of solar power in~$2019$ exceeding $921$~GWh \cite{CAISO2017k}.

\section{Methodology}
\label{sec:methodology}
This section outlines the general framework that is used for risk-based adequacy and capacity-value assessments in systems with substantial VG penetrations. Most of the material is applicable equally to all VG. Thus, this material seldom makes specific reference to solar power. Specific consideration of energy storage is beyond the scope of this paper. Thus, energy storage is not discussed except in Section~\ref{sec:methodology:hybrid}.% on sites with integrated PV and storage.
%\textcolor{purple}{[SA: it may be useful to cite some papers that clarify what the time-collapsed model is?] }

\subsection{Probability Background}
\label{sec:methodology:probability}
In adequacy assessment, we are interested in the values of available conventional capacity, $X_t$, available VG capacity, $Y_t$, and demand, $D_t$, during multiple points in time, which are indexed by $t$. Let the (random) vector, $S_t=(X_t,Y_t,D_t)$, denote the system state at $t=1, \dots, n$ within the period that is under study. The system margin, $Z_t=X_t+Y_t-D_t$, is a function of $S_t$. A full probability model for the system would be sequential, describing $S_t$ as a stochastic process over the entire time period. Such a stochastic process is needed to calculate some risk metrics, \eg, frequency and duration indices, or the distribution of total energy unserved across the period under study.

However, some quantities, such as loss-of-load expectation (LOLE), which is defined as:
\begin{equation}
    [\text{LOLE}] = \sum_{t=1}^n \prob{Z_t<0},
\end{equation}
may be defined in terms of the marginal distributions of $S_t$ integrated over time. %\alw{looking at this again, I don't know if saying "defined in terms of the marginal distributions of $S_t$" is confusing as I don't think you can get the marginal distribution of $Z_t$ from knowing the marginal distributions of the individual components of $S_t$ unless they are independent? What about switching $S_t$ here with $Z_t$?} \sz{While Amy's assertion is correct, I don't think there is any problem here, as the distribution of the "time-collapsed'' r.v. $S$ is a 3-dimensional joint distribution by definition, i.e.\ more than just the distributions of $X$, $Y$ and $D$. Hence we have $P(Z<0)=P(X+Y-D<0)=1/n\sum_{t=1}^nP(X_t+Y_t-D_t<0)=1/n\sum_{t=1}^nP(Z_t<0)$},
%\cjd{Need to resolve question between Stan and Amy, now commented out.} \sz{I think this is now resolved}, 
LOLE may be specified equivalently in terms of a simpler \emph{time-collapsed} or \emph{snapshot} model with a time-independent state vector, $S=(X,Y,D)$, the distribution of which is specified by:
\begin{equation}
\label{equ:methodology:probability:lole:timeIndep}
    \prob{S \in A} = \frac{1}{n} \sum_{t=1}^n \prob{S_t \in A},
\end{equation}
for any event, $A$. In~(\ref{equ:methodology:probability:lole:timeIndep}) the distribution of the state vector, $S$, is the same as that of state vector, $S_t$, sampled at a uniformly randomly chosen point in time. The specification in~(\ref{equ:methodology:probability:lole:timeIndep}) is helpful for some computational or theoretical analyses. Using~(\ref{equ:methodology:probability:lole:timeIndep}), LOLE is given as $(\Delta t) \prob{Z<0}$, and expected energy unserved as $(\Delta t) \E{\max \{-Z,0\}}$, where $Z=X+Y-D$ and $\Delta t$ is the length of the period under study. The distribution of $S$ typically is estimated from the empirical distribution of observations of $S_t$. Thus, the time-collapsed model is used almost always in adequacy studies that measure risk using quantities, such as LOLE, which do not require a full sequential model.

\subsection{Statistical Estimation}
In the use of probabilistic and statistical concepts such as independence or correlation, it is essential to be clear as to which of the sequential and time-collapsed models these refer. For example, suppose $Y_t$ is available solar power at time~$t$ and that at any given time, $t$, the random variables, $Y_t$ and $D_t$, are independent (neither being informative about the other given the  knowledge at time~$t$). Because daily minimum demand usually occurs overnight when it is dark, within the \emph{time-collapsed} model the lowest values of $D$ are associated with zero values of $Y$, introducing substantial probabilistic dependence between these two time-collapsed random variables.

In reality, even conditional on information at time~$t$, there is typically still some dependence between variable generation, $Y_t$, and demand, $D_t$, due to the existence of unmodelled weather effects, which influence both $Y_t$ and $D_t$. This modifies the dependence between the corresponding time-collapsed random variables, $Y$ and $D$.
%\sht{Also, to clarify this concept for outsiders, I suggest turning the argument in this paragraph around: D and Y are dependent, largely due to (hidden) time and weather. The variables $D_t$ and $Y_t$ are conditioned on specific times, reducing their dependence. Nevertheless, the weather (not explicitly modelled) remains a factor that affects both.} \cjd{Stan, any thoughts?} \sz{I have done some considerable rewriting to take account -- I hope -- of Simon's very good suggestion.}

If dependence between VG output and demand is considered in a time-collapsed model, often this is done using a `hindcast' approach, in which the empirical historical distribution of VG-output/demand pairs, $(y_\tau,d_\tau)$, 
%(see the comment at the end of Section~\ref{sec:methodology:probability}) 
is used as the predictive joint distribution of $(Y,D)$. 
%\ils{This feels cryptic - I can't find a comment or footnote a the end of III-A.}
The random variable, $X$, usually is assumed independent of the pair, $(Y,D)$, with a distribution estimated from an appropriate model. Then:
\begin{equation}
    [\text{LOLE}] = \frac{\Delta}{N} \sum_\tau \prob{X+y_\tau < d_\tau},
\end{equation}
where $\Delta$ is the length of a time step, $N$ is the number of historic years of data, and the sum is over historic times, $\tau$.

%An issue with the hindcast approach is that 
Inevitably, there are limited relevant data in the hindcast approach for estimation of the empirical distribution at times of high demand and low VG output, which dominate the estimates of risk measures. This can be dealt with by using statistical extreme-value theory to smooth the extremes of a dataset \cite{Wilson2019}. To the best of our knowledge, the only works using more sophisticated direct joint modelling of the relationship between VG output and demand in a time-collapsed model are the work of Wilson \etal\ \cite{wilson2018a} (which uses temperature as an explanatory variable for both wind and demand, and invokes independence of wind and demand conditional on temperature and on time of day, week, and year); and the work of Gao and Gorinevsky \cite{Gao2018} (which uses quantile regression to model explicitly the distribution of wind conditional on demand).
% \ils{This sentence is very tough to read - two long bits in brackets - and I think 'are' before [48] is a mistake} \cjd{I have altered punctuation to make this clearer}

Studies that consider estimation of the uncertainty that arises from the use of limited numbers of years of data typically assume that a result derived from the longest available dataset is `the truth' \cite{hasche,MadaeniSioshansiDenholm2012b}. However, this is not fully satisfactory, as the result may be driven by a small number of historic weather systems, and there may be a tendency for extreme peaks to cluster in neighbouring years, reducing further the number of fully independent datapoints. Some discussion of this is provided in the literature \cite{wilson2018a,Wilson2019}, although more work in this area and on the consequences for decision support is required.
%\ils{'are is' doesn't make sense.  What should this sentence say?}

Most studies using a \emph{sequential} model assume that VG output and demand may be modelled as independent processes within the season under study \cite{billinton2011a,troffaes2015a}. In reality, as discussed above, some dependence between these processes may be introduced by the variability of the weather. There is little research on multivariate-stochastic-process modelling of VG output and demand for adequacy assessment \cite{wilson2016a,antarescore}. %\sht{the Antares reference is to the tool itself. There is also a 2011 Trondheim PowerTech paper by Doquet \etal that we could use instead.} \cjd{Please provide a citation.} \sht{A time series generator is described (at a high level) in 10.1109/PTC.2011.6019444, but I checked and it only considers multivariate wind series; apparently no correlation with demand. Perhaps they have built in such a feature in recent years.}

\subsection{Capacity-Value Metrics}
\label{sec:methodology:metrics}
Capacity-value metrics are used commonly to visualise the contribution of VG (or other resources) in adequacy studies \cite{Keane2011a}. For instance, in the time-collapsed model and with respect to the loss of load probability (LOLP) risk index, the effective load-carrying capability (ELCC) of a resource, $Y$, when added to a background, $M$, is given by the solution of:
\begin{equation}
\label{equ:methodology:metrics:elcc}
    \prob{M<0} = \prob{M+Y < [\text{ELCC}]_{Y,M}},
\end{equation}
and the equivalent firm capacity (EFC) is given by solving:
\begin{equation}
\label{equ:methodology:metrics:efc}
    \prob{M+Y<0} = \prob{M + [\text{EFC}]_{Y,M} < 0}.
\end{equation}
These capacity-value metrics are functions of the chosen risk metric and the background, $M$, to which it is added, as well as of the additional capacity. Thus, it is incorrect to refer to the capacity value of $Y$ without that \textit{caveat}, or to use a single capacity-value figure across multiple circumstances \cite{Zachary2019}. This nuance is particularly important in capacity-market applications. Such capacity-value metrics are also non-additive, \ie, the ELCC (or EFC) of an addition, $Y_1+Y_2$, typically will not equal the sum of the ELCCs (of EFCs) of $Y_1$ and $Y_2$ added to the same background.

As is clear from~(\ref{equ:methodology:metrics:elcc}) and~(\ref{equ:methodology:metrics:efc}), when adding a single relatively small resource to the background of a much larger system, ELCC and EFC take very similar values. This similarity applies when calculating the marginal capacity value of a single unit in a capacity market. In other applications, it might be of interest to calculate the capacity value of an entire fleet of wind or solar generation when added to the background of the other resource and demand. In such cases, ELCC and EFC may take different values and it is necessary to consider which capacity value metric is appropriate. ELCC is used most commonly, however it is not always clear whether this choice is considered carefully with respect to the specific application.

Various special cases (\eg, small $Y$ and exponentially distributed $X$) are surveyed by Dent and Zachary \cite{Dent2014a}, building on earlier work \cite{garver1966a,dragoon2006a,dannunzio2008b,zachary2012a}. These cases are helpful in understanding what is driving the results of capacity-value calculations. Computation is usually sufficiently straightforward that these special cases are not needed typically for model tractability.

\subsection{Including VG in Capacity-Remuneration Mechanisms}
\label{sec:methodology:capMech}
\begin{figure}
\centering
\includegraphics[width=\columnwidth]{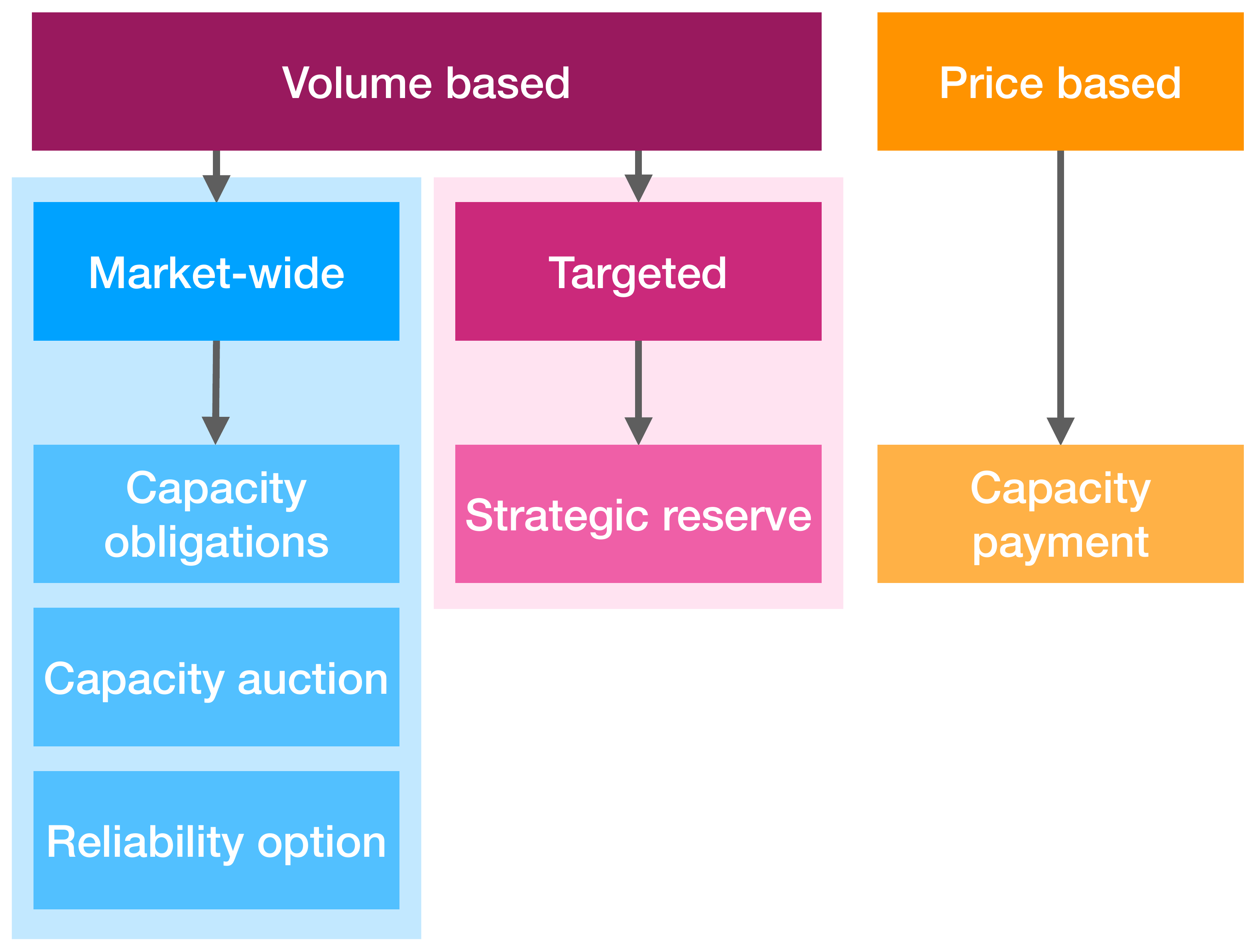}
\caption{Taxonomy of capacity-remuneration mechanisms \cite{capacity}.} %\sjp{I dropped in a reworked version of this, CRMv2.pdf. Old figure still here. Happy to modify further if it seems useful.}}
\label{fig:methodology:capMech:taxonomy}
\end{figure}

Capacity-remuneration mechanisms (CRMs) incentivise the presence of an appropriate level of generation and equivalent capacity for resource-adequacy purposes. They take a range of forms, with a useful taxonomy that is provided by Agency for Cooperation of Energy Regulators (ACER) \cite{capacity} and summarized in Fig.~\ref{fig:methodology:capMech:taxonomy}. Further detailed surveys of CRMs may be found in other works \cite{ferc,creg,doorman2016}, with Table~$0.1$ in the latter providing a more granular taxonomy than that of ACER. For crediting VG in CRMs, appropriate modelling of the adequacy contribution of the resource is needed. This applies similarly to all volume-based mechanisms, and in a different manner to price-based mechanisms. Thus, this section describes the theory behind volume- and price-based CRMs, particularly the role of capacity-value metrics in including offers from VG.

\subsubsection{Volume-Based CRMs}
Here a central authority defines a volume of capacity to procure, \eg, based on a target risk level or a cost-benefit analysis. Then, typically an auction is held to determine the units that are selected and the capacity price.

There is a standard theory for capacity procurement in volume-based markets, in which all offers are from resources equivalent to conventional generation \cite{Zachary2019}. Suppose that (to a good approximation) adding or subtracting a limited capacity of conventional resource shifts the distribution of the margin, $Z$, with changes in the shape or width of that distribution being a lower-order effect. Then, it is possible to define the volume of capacity in terms of expected available capacity, with the product offered by an individual unit being its expected available capacity. Units are added in ascending order of their ratio of offer price to expected capacity, until the sum of their expected available capacities equals the target. This is referred to then as an auction with expected available capacity (sometimes referred to as `de-rated capacity') as a `simple additive commodity'. Without significant additional complication, the fixed capacity target could be replaced by a demand curve, implying that at a higher auction price the amount procured will be lower.

%If all resource were firm (i.e. delivered a given rated capacity with probability 1) then it would be a simple matter to set a risk target with respect to a chosen index (e.g. LOLE) and add units in order of increasing (capacity / offer price) until the risk target is met. This same theory applies in practice to a system with only conventional generation, as to a good approximation adding or subtracting a small number of units shifts the probability distribution of margin (changes in its width or shape being a lower order effect) -- thus expected or `de-rated' capacity may be used as an additive commodity in an auction.

The assumptions that are required to run an auction with an additive commodity do not hold when non-conventional resources, such as VG or energy storage, participate in the market. 
%\sht{I wonder: do they hold for conventional generators? Consider a mix of small and large generating units. The same caveats would apply.} \cjd{I have covered this above to some extent. I can ask Nestor to check out the Irish case.} 
Instead, the above mechanism may be generalised by adding units in ascending order of the ratio of offer price to the marginal EFC against the background of the finally accepted set of resources, until a specified risk target is reached. Crucially, however, the final accepted set of resources cannot be known \textit{ex ante}. Thus, it is necessary to perform an iterative process of running the auction and recalculating EFCs with the latest auction outcome, until convergence is obtained \cite{Zachary2019}. This is in contrast to how quantity-based capacity markets operate currently, wherein all bidders submit price/quantity offers that are based on their (possibly de-rated) capacity, which is determined \textit{ex ante}. Therefore, quantity-based CRMs (as structured currently) cannot consider contributions of all types of resource on an equal basis.

Volume-based CRMs typically require specification of a penalty if a contracted resource cannot deliver when required. One specific form of penalty is a reliability option (RO) \cite{vazquez2002market}, which is a one-way contract for differences against the energy-market price. Whenever the market price rises above a specified level, any firms that hold a RO are required to pay the difference between the strike price and the market price to the system operator. VG can face significant risk in taking on such contracts, due to its uncertain and variable output. However, this is not discussed in detail here as penalty regimes are a separate matter from capacity value and procurement.

%Reliability options are a form of volume-based mechanism, in which the penalty for non-delivery takes the form of a reliability option (RO). A RO is a one-way contract for differences against the energy market price; whenever the market price rises above a specified level, any firms that hold a RO are required to pay the difference between the strike price and the market price to the TSO.
%Reliability options were originally proposed by \cite{vazquez2002market}. The direct link to the energy market means that generators that cannot reliably provide electricity during any given period, face  significant potential exposure to energy market prices. This means that VG in particular faces extra risk in taking on these contracts.  \sz{It is not described how this forms part of a mechanism for obtaining sufficient capacity.} \cjd{I need to discuss how we edit this section with Muireann}

\subsubsection{Price-Based CRMs}
Under price-based CRMs, the regulator or system operator determines the total remuneration for capacity, and how this is assigned \textit{ex post} to resources according to their performance. The total capacity investment is a market outcome, based on incentives provided by the CRM and other sources of income.

Total remuneration typically is calculated as the product of a volume element (the total generation capacity that is required to ensure system adequacy) and price element. The volume element is calculated similarly to the capacity target for a volume-based mechanism, and 
%,  based on projected peak demand, interconnector flows\footnote{Incorrect assumptions surrounding interconnector flows can lead to over- or under-procurement of capacity. For a full description see \cite{newbery2014final}. \cjd{Maybe move volume based mechanisms first?}} and also on forced and scheduled outage rates of conventional generation. 
is multiplied by a specified per-MW cost of new entry to give the total remuneration. Variants include pre-$2001$ England and Wales, wherein there was no fixed capacity payment. Instead, for each time the total payment was the product of day-ahead LOLP and a specified value of lost load \cite{newbery}.

Price-based mechanisms do not require the use of a de-rating factor or capacity value in a capacity auction, as the outcome of the generator availability is used to distribute the revenues. Thus, the complications surrounding \textit{ex ante} assignment of capacity values do not arise. However, this means that resources are rewarded implicitly on the basis of some form of mean output, which may not reflect well a resource's contribution within an \textit{ex ante} risk calculation. This is particularly problematic for VG, the contribution of which within probabilistic risk calculations can be much less than that of firm capacity equal to its mean output.

\subsection{Generation-Expansion Models}
Several works embed adequacy risk calculations in generation-expansion optimization models \cite{munoz2015a,sigrin2014a,bothwell2017}. These works minimise the cost of capital investment, unserved energy, and (possibly) operations. Typically, unserved-energy costs are included through a hindcast risk calculation using multiple years of demand and VG-output data. To give a linear optimization model, it is necessary to simplify representation of conventional generators, \eg, assuming that conventional-plant availability is deterministic and equal to its mean.

Bothwell and Hobbs \cite{bothwell2017} assess social-welfare losses if VG capacity is credited inappropriately and express the value of additional VG in terms of its marginal EFC at an economic optimum. They do not provide, however, a practical scheme for operating a CRM with both VG and conventional generation.

In energy-system models with wider scope, \eg, The Integrated MARKAL-EFOM System (TIMES), security of electricity supply is represented typically \textit{via} a target de-rated margin of installed capacity over peak demand \cite{price}, as embedding any kind of risk calculation would be too computationally expensive.

\subsection{Hybrid VG and Energy Storage}
\label{sec:methodology:hybrid}
At a system level, energy storage can enhance the capacity value of VG \cite{Stenclik2018}. Here we consider integration of energy storage with VG at a single site (\ie, with a single grid connection). Such energy storage can be inherent in the VG, as for CSP plants \cite{pfenninger2014a,madaeni2013b}, or dedicated energy storage that is co-located with VG, as in grid-connected microgrids \cite{Mitra2012}. Typically, integrated energy storage can recharge only from the associated VG resource (\eg, heat from irradiance in the case of CSP), and not directly from the grid \cite{rustomji2016}. Thus, capacity value can be computed only for the integrated system.

%It is well-established that at a system level, the presence of storage enhances the capacity value of variable generation. This has, for example, been investigated for the case of solar PV and battery storage in \cite{Stenclik2018}. In this section we consider the effects of integrating storage with VG at the level of a single site (i.e. with a single grid connection). Such storage can be inherent in the energy production process, as is the case for concentrating solar power (CSP) \cite{pfenninger2014a,madaeni2013a}, or a dedicated storage facility can be co-located with VG, as is the case of grid-connected microgrids \cite{Mitra2012}. Integrated storage can typically only recharge using the associated variable generation resource (e.g. heat from irradiance in the case of CSP), and not directly from the grid \cite{rustomji2016}. As a result, a capacity value can only be computed for the integrated system.

Examples of such capacity-value estimations for CSP include the work of Madaeni \etal\ \cite{madaeni2013b}, which uses a capacity-value approximation that is based on the $10$~highest-LOLP hours of each year. They conclude that increased energy-storage capacity increases capacity value and reduces its interannual variation. Usaola \cite{usaola2013a} study a CSP plant with deterministic dispatch using a time-sequential Monte Carlo calculation and obtain qualitatively similar results, with differences arising from sizing of the CSP plant and different generation and demand statistics. Mills and Rodriguez \cite{mills2019} consider a looser form of coupling, wherein PV that is co-located with batteries share inverters, necessitating an integrated assessment.

%Examples of such studies include Madaeni \emph{et al}, who consider the capacity value of CSP plants with integrated storage \cite{madaeni2013a}, computed using a capacity value approximation based on the 10 hours of each year with the highest LOLPs. They conclude that increased storage capacity not only increases the capacity value, but also reduces the year-on-year variation of its effective annual capacity value. Usaola \cite{usaola2013a} studies a CSP plant with a deterministic dispatch strategy using a time-sequential Monte Carlo method and obtained qualitatively similar results, albeit with smaller magnitude due to a larger relative size of the CSP plant, and reduced coincidence of generation and demand. Mills and Rodriguez \cite{mills2019} consider a looser form of coupling where co-located PV and batteries share inverters, necessitating an integrated assessment.

On the other hand, if VG and energy storage can be operated independently (\eg, a battery with a separate inverter), the capacity value of the integrated system may be calculated as the sum of capacity values of its constituent components, \emph{if} two conditions are satisfied. First, the contribution of the VG and energy storage must be small with respect to the total system size, so that their capacity values are marginal \cite{Zachary2019}. Second, each constituent capacity-value calculation must account for the ability to re-dispatch existing generation and energy storage. The difference of this \emph{integrated capacity value} and a simple dispatch adjustment can be very substantial---up to an order of magnitude for a combination of pumped hydroelectric energy storage and solar \cite{Karier2017,Byers2019}.

\section{Survey of Current Practice}
\label{sec:survey}
This section reviews the literature to illustrate points made earlier. It does not attempt an exhaustive literature survey of practice, as in Doorman \etal\ \cite{doorman2016} and S\"{o}der \etal\ \cite{Soder2019}, which are referenced as relevant. The number of individual works that are cited in this section is relatively small, as many studies use similar methodologies. One limitation of many broader surveys is that they do not provide our critical discussion of technical modeling approaches.

%Useful broad surveys of industry practice include \cite{doorman2016} (the report of a Cigre working group, providing a comprehensive discussion of issues in capacity market design along with detailed case studies), and \cite{Soder2019} (providing descriptions of wind generation adequacy calculation practices in a large number of different systems). Neither, however, provides this paper's critical discussion of technical modelling approaches.

%may be found for instance in [ZhouFrew,SarahHamid] (on capacity value methodology) and [Batlle] (on capacity market designs).
%\textcolor{purple}{[SA:In this TF paper, we sometimes refer to capacity value as capacity credit eg Section IV A – we might want to be consistent.]} \cjd{I have now edited for consistency}

\subsection{Recent Methodology-Related Research}
As described in Section~\ref{sec:methodology}, if a statistical relationship between VG output and demand is taken into account, this is done typically through the `hindcast' approach. We note examples of formative works taking such an approach with wind \cite{Keane2011a,Soder2019} and solar generation \cite{gami2017a,abdel2017} generation. Several studies review the variants of methodology that are used in different studies or the consequences of different approaches for numerical results. Mills and Wiser \cite{mills2012b} provide a list of the capacity-value approaches that are used in different utilities for planning purposes. Madaeni \etal\ \cite{madaeni2012comparison} use the western United States as a case study, Zhou \etal\ \cite{zhou2018valuing} emphasise the impacts of mis-estimating capacity value, and Awara \etal\ \cite{Awara2018} survey the impact on calculation results of making different modelling decisions.

Other recent research considers associated data issues. Gami \etal\ \cite{gami2017a} examine consequences for calculation results of input data resolutions such as temporal resolution and ambiguity over definitions of data fields in recording PV output. Madaeni \etal\ \cite{madaeni2013b} use hindcast to compare how different approximations to the full risk calculation affect LOLE-based ELCC results. Abdel-Karim \etal\ \cite{abdel2017} demonstrate carefully how issues in data rounding affect comparison of results from different codes, in the context of using the hindcast approach on the IEEE Reliability Test System.

%\textcolor{purple}{[SA:Alberta’s capacity market design was finalized, but did not go into effect because of the change in government (worth mentioning in Section IV B)]}
%\cjd{These last two paras might shift to next subsection if we make IV-A relevant research, and IV-B industry practice}

\subsection{Capacity Markets}
%\todo[inline]{Do we need to mention the recent reviews \cite{Hoschle2016} amd \cite{Soder2019}? I have uploaded the CIGRE doc to overleaf, in case people don't have access.} \cjd{Added to the start of Section IV}

%The calculations of risk and capacity values described above are applied in various ways in modern capacity markets. Referring to the ACER taxonomy in figure \ref{fig:methodology:capMech:taxonomy} above, variable renewable generation generally does not participate in targeted mechanisms such as strategic reserves. Risk calculations are therefore relevant only at the stage of calculating the quantum of reserves to procure, but not in determining the appropriate remuneration for these reserves. Similarly, price based mechanisms, as outlined above, do not require an \emph{ex ante} calculation of the capacity contribution of each resource, including variable renewable generation resources. Thus the \emph{ex ante} calculation of capacity values is required for market-wide quantity-based capacity mechanisms such as capacity auctions.

Capacity-value metrics for VG are of most relevance in volume-based CRMs: renewables often do not participate in strategic reserve/targeted mechanisms. Price-based CRMs do not require assigning a capacity value \emph{ex ante} (\cf\ Section~\ref{sec:methodology:metrics} and examples such as the Nordic system \cite{Soder2019}).

In volume-based CRMs, the most common method of accounting for the adequacy contribution of different technologies is application of a de-rating factor. Thus, a unit is compensated for only a portion of its nameplate capacity in auction processes and in consequent payments, to account for its estimated statistical availability properties. 
%to account for forced and scheduled outages or unavailability during times of zero renewable output. 
Mean availability is used typically for conventional generation.

Applying an appropriate de-rating factor to VG is challenging, however, as we discuss previously. A range of modelling approaches for resource-adequacy assessments, partly based on the characteristics of the relevant power system, can be used. Bothwell and Hobbs \cite{bothwell2017} and S\"{o}der \etal\ \cite{Soder2019} include surveys of current practice in North America and Europe, with the latter examining the case of wind generation only but providing a survey of a much larger number of systems.

Table~$3$ in the work of S\"{o}der \etal\ \cite{Soder2019} summarises the methods that are used to determine capacity value of wind in the systems that are surveyed. Where wind is eligible for capacity payments, a risk-based capacity-value metric is used typically, \eg, marginal EFC in Great Britain, average EFC in Italy, and marginal ELCC in Ireland. Some systems, particularly those that rely on strategic reserves, such as the Nordics, preclude renewables from receiving payments at all. Great Britain permits wind generators to receive a capacity payment if they are not in receipt of low-carbon support, which in practice means that most wind farms do not participate. The Irish and Italian systems allow all renewable projects to participate in capacity auctions. However, to date, renewable projects represent only a tiny proportion of successful offers in Ireland and Italy.

%Other examples include a fully worked proposal for AESO 
%\todo[inline]{HZ [AESO's capacity market is cancelled]} 
%(the AESO capacity market was not in the end deployed), which used the mean output of the solar farm during the 250 hours of lowest (rather than peak) historic supply-demand margin from the previous 5 years  \cite{AESO}. This is because AESO sees its highest risk hours when plant is on maintenance. In such an example, one should however take great care in statistical estimation and risk modelling if there is a possibility of rescheduling maintenance when margin is very tight.

%Other operators using means (or other related approaches, often collectively referred to as `approximation' approaches) for solar power include MISO \cite{MISO_BPM_011} and ISO-NY \cite{NYISO}. However, PJM is proposing to change its current solar capacity value calculation to an ELCC approach as described in Section III \cite{PJM_ELCC}. Likewise, MISO intends to explore using ELCC  for solar resources when more data is available \cite{MISO_BPM_011}, as it already does for wind. 

%(eg, Britain, Italy) the EFC is used to determine the payment due to wind producers (although in the case of Britain the marginal EFC is used while in Italy the average EFC is calculated). In the Irish market, the ELCC of wind is used, under through a least-worst regrets approach. However it is again the average rather than the marginal ELCC that is calculated.

From a risk-modelling perspective, there are different contexts in which it may be necessary to consider VG within capacity auctions. Clearly, in systems in which VG receives a capacity payment on the basis of a risk-based capacity-value metric, it must be included in the risk calculations. There are other examples in which VG does not receive capacity payments, but is included in the risk modelling which underpins the capacity market, \eg,
%, Germany, where wind in the north leads to the need for grid reserve in the south, or 
in Finland, where wind can reduce the need for strategic reserves. In other markets (\eg, Sweden), wind is excluded explicitly, which potentially could lead to over-procurement of other capacity.

%finds that the current practice surveyed fails to credit wind and solar appropriately: the calculation of the derating factor in the markets surveyed mostly depends on mean or median load factors over a defined subset of high demand hours (ERCOT, IESO, ISO-NE, NYISO, PJM), or a higher percentile of the distribution of output in a given month (CAISO). Only MISO used a risk-based capacity value (ELCC calculated separately for conditions in each historic year, and averaged over 10 historic years). While peak period means may be relevant in some circumstances (e.g. very small VG capacities), if they are to be used it is necessary to assess carefully whether they represent appropriately the contribution of the VG to mitigating adequacy risk.

Other systems use a summary statistic of an estimated probability distribution of available resource to represent the contribution of VG in capacity markets or policy-facing resource-adequacy studies. For instance, PJM uses the mean conditional on summer-peak hours, Texas uses mean from highest-load hours during the previous $10$~years, Spain uses the lower fifth quantile of the distribution, and a system that was proposed (but never implemented) in Alberta uses $250$~hours of lowest historic margin during the last five years, which accounts for significant risk contribution in the maintenance season. All of these approaches credit VG on the basis of its own properties, \ie, in contrast with a risk-based approach, not on how its properties affect the risk level in the system as a whole. This property of these approach has potential serious consequences when VG penetration is very high, as it is in Texas. However, these approaches may be more appropriate at very low penetrations of VG, which can be checked on a case-by-case basis. Bothwell and Hobbs \cite{bothwell2017} examine the economic consequences of using alternatives to an appropriate risk-based capacity credit (\eg, techniques that are employed in ERCOT, IESO, ISO New England, PJM, and California).

It is not clear in all cases whether historic metered output is used, or whether historic meteorological data are used in combination with a future scenario of installed VG capacity. The former has the advantage of being based on actual historical performance, whereas the latter often is preferable as it permits consideration of newer or future sites where there is little or no metered historic record.

%\cjd{I have not edited this para yet, pending additional interesting examples.} The methodology employed in the Irish Integrated Single Electricity Market (I-SEM) is essentially a calculation of average ELCC relative to the LOLE target \cite{cru1}. In Great Britain, derating factors are calculated using incremental EFCs rather than average, leading to very low derating factors for wind and solar \cite{emr1}. One thing these methods have in common is that they rely on the use of historical data to calculate a de-rating factor, and in the case of those practices surveyed in \cite{bothwell2017} at least, fail to credit wind and solar appropriately.

\section{Conclusions}
\label{sec:concl}
This paper reviews methods that are used for adequacy risk assessment considering solar power and other VG technologies, and for assessing the capacity value of VG installations. This includes the spatial and temporal properties of solar output, solar-design considerations, methods for capacity-value assessment,% (including relevant statistical considerations)
and including VG in CRMs. Our survey of current practice reveals broad heterogeneity, confirming that a review paper of this type is warranted.

Although there is a growing literature on reliability assessment and capacity value considering solar and other VG, several outstanding issues call for additional research. While considerable advances have been made in resource assessment of solar and wind power, there is little work on building error models quantifying the consequences of uncertainty in reconstruction of historic resources. Further statistical work on resource-adequacy assessment is needed. This includes work on non-sequential approaches beyond hindcast and joint VG/demand modelling for sequential models and on use of these more advanced approaches in practical circumstances. The overall emphasis should be on how these various developments could improve decision analysis. Finally, there is limited understanding of how to operate capacity markets on a technology-neutral basis with a full range of resources, including conventional plant, VG, energy storage, and other emerging resources.

%Capacity markets should consider the interplay between resource availability and dispatch and how different VG resources (\eg, PV versus CSP) should be incorporated. Many existing methodologies rely on hindcasting. A better undersstanding of the limitations of hindcast and understanding the impacts of limited data on uncertainty in estimating capacity values is needed.

%\section*{Acknowledgments}

\bibliographystyle{IEEEtran}
\bibliography{SolarTFpaper}
\end{document}